\documentclass[a4paper, 11pt]{article}
\usepackage{fullpage} 
\usepackage{amsmath}
\usepackage{amsfonts}
\usepackage{graphicx}
\usepackage{amssymb}
\usepackage{color}
\usepackage[usenames,dvipsnames,svgnames,table]{xcolor}
\usepackage{soul}
\usepackage{cite}
\usepackage{flushend}
\usepackage{algorithm}
\usepackage[noend]{algpseudocode}
\makeatletter
\def\BState{\State\hskip-\ALG@thistlm}
\makeatother

\DeclareMathOperator*{\argmin}{argmin}

\usepackage[utf8]{inputenc}
\usepackage{graphicx}
\title{Market based embedded Real Time Operation for Distributed Resources and Flexibility} 
\date{May 5, 2017}
\author{Bertrand Travacca, Greg Rybka, Kenji Shiraishi}
\begin{document}
\maketitle

\section{Abstract}
We build upon previous work out of UC Berkeley's energy, controls, and applications laboratory (eCal) that developed a model for price prediction of the energy day-ahead market (DAM) and a stochastic load scheduling for distributed energy resources (DER) with DAM based objective\cite{Travacca}. In similar fashion to the work of Travacca et al., in this project we take the standpoint of a DER aggregator pooling a large number of electricity consumers - each of which have an electric vehicle and solar PV panels - to bid their pooled energy resources into the electricity markets. The primary contribution of this project is the optimization of an aggregated load schedule for participation in the California Independent System Operator (CAISO) real time (15-minute) electricity market. The goal of the aggregator is to optimally manage its pool of resources, particularly the flexible resources, in order to minimize its cost in the real time market. We achieves this through the use of a model predictive control scheme. A critical difference between the prior work in \cite{Travacca} is that the structure of the optimization problem is drastically different. Based upon our review of the current and public literature, no similar approaches exist. The main objective of this project were building methods. Nevertheless, to illustrate a simulation with 100 prosumers was realized. The results should therefore be taken with a grain of salt. We find that the Real Time operation does not substantially decrease or increase the total cost the aggregator faces in the RT market, but this is probably due to parameters that need further tuning and data, that need better processing. 

\section{Introduction}
\subsection{Motivation and background}
As a global leader in climate change policy, California has one of the highest renewable energy target in the world; 50\% of electricity supply will be from renewable electricity sources (RES) by 2030.
\footnote{In early 2017, California Senate leader Kevin de Le\'{o}n put forth a bill that would mandate the State to use 100\% renewable power by 2045.  So this issue of addressing intermittent resources is unlikely to be resolved through lower legislated goals, but rather increased goals only tightening existing system constraints.} 
Most of the RES that California uses are variable and intermittent such as wind and solar power. The need to ramp up or ramp down controllable, often non-renewable, generation sources due to the variability of the RES is a current and increasing challenge for power system operators and policy makers.\\ 

Historically, supply-side resources have followed the load.  Load, in general, has been considered to be inelastic, as the prices that are observed by retail customers do not correspond to wholesale electricity prices.  With the increase in variable supply resources there is increasing interest in the use of demand-side resources to participate in the wholesale markets as a mechanism to resolve some of the supply-demand imbalances that could otherwise occur.\\  

In this context, both demand response and energy storage are considered possible mitigating measures and technologies that can provide flexibility to the grid.  This helps to address problems associated with the famous "duck curve" and the power ramping problem in the late afternoon.  The potential net benefits of the use of demand response and energy storage have not been well studied based on real data.\\

The California Independent System Operator (CAISO) operates three distinct markets, a day-ahead market, real-time market, and ancillary services.  Ancillary services, such as congestion revenue rights and convergence bidding, is also a set of markets that the ISO operates though they will not be addressed in this analysis.\\

The day-ahead market (DAM) has three separate processes.  The first is to assess whether the bidders may be able to exert market power, the second is to forecast the level of supply needed to meet the demand, finally the additional plants that must be ready to generate are determined.  Bids and schedules can be submitted up to seven days in advance, but the market closes at noon the day before and the results are distributed at 1pm.  As a part of this market, bilateral trading occurs weeks in advance.  These are then scheduled at noon, prior to the ISO publishing the results of the optimization being run.  About 70\% of energy in CAISO goes through the market as self schedule or as a price taker.\\

The real-time market (RTM) offers an opportunity for the CAISO to procure supply closer to when the demand occurs, thus the forecast is less uncertain.  The market opens at 1pm the day prior to the trading day (i.e. after the DAM results are published) and closes 75 minutes prior to each trading hour, with the results being distributed 45 minutes prior to the hour.  The dispatch occurs in 15 and 5 minute intervals, depending on the plant.

\subsection{Relevant Literature}
A growing body of literature addresses optimal plug-in electric vehicles (PEV) population charging and residential demand response \cite{RTM stochastic optimization 1}, \cite{RTM stochastic optimization 2}, \cite{centralized1}, \cite{centralized2}, \cite{centralized3}. Within the literature that we examined, we found studies that explored either optimization in the DAM, the RTM, or TOU (Time-of Use scheduling).\\
\\
Market bidding strategies and market uncertainty for aggregated PEVs have been studied in several research groups and results can be found in the following publications: \cite{marketObjective1}, \cite{marketObjective2},\cite{marketObjective3}. These studies analyzed both bidding in the DAM and RTM separately.
\\
In this paper, we construct a tailored optimization method for real time operation of  electricity resources in the Real Time Market using our aggregated resources. It is important to note, that it is no longer simple scheduling as in \cite{Travacca}.

\subsection{Notation and Nomenclature} \label{notation} 
For $(x,y) \in \mathbb{R}^d$, $<x,y>=x^Ty$ refers to the euclidean scalar product of $x$ and $y$, $ \|.\|_2^2$ refers to the corresponding euclidean norm. $x \leq y$ refers to element-wise inequality. $\odot$ denotes the element-wise vector product (a.k.a hadamard product). For a vector $v\in \mathbb{R}^d$, $v(a:b)\in \mathbb{R}^{b-a+1}$, with $a$ and $b$ integers, denotes the vector consisting of the $v_j$, $j\in\{a,...b\}$. \\
The following notation is used in this paper.  Uppercase letters refer to variables with units of power ($kW$) while lower case letters refer to variables with units of energy ($kWh$). Symbol $x^t$ refers to the value taken by variable $x$ at time $t$. In the absence of the exponent we will consider the variable $x$ as a vector $\in \mathbb{R}^{24}$. Symbol $x_i$ refers to a local variable $i\in \{1,...N\}$. Finally, $\overline{x} $ (respectively $\underline{x} $) refers to an upper (lower) bound of the variable $x$. For more clarity, current decision variables are highlighted in \textcolor{orange}{orange}.

\renewcommand{\arraystretch}{1.8}
\begin{table}[H]
\caption{Nomenclature}
\label{nomenclature}
\begin{center}
\begin{tabular}{c l}
\hline \hline
$N$ & Number of prosumers \\
$\Delta t$ & Time-step for Real Time Market: 15 min\\
$T$ & Time-Horizon (hours)\\
$T_H$ & Model Predictive Control Time-Horizon (hours)\\
$\lambda_{DA},\lambda_{RT}$& respectively DA and RT risk aversion parameters\\
& All of the following variables are $\in \mathbb{R}^{T}$\\
$p_{DA}$ & Day-Ahead Market price\\
$L_i$ & Uncontrollable residential load of prosumer $i$\\
$S_i$ & Solar PV production of prosumer $i$ \\
\textcolor{orange}{$EV_i$}& Day Ahead Charging rate of PEV  $i$\\
\textcolor{orange}{$G_i$}& Day Ahead Power imported from the grid for prosumer $i$\\
& All of the following variables are $\in \mathbb{R}^{4T}$\\
$p_{RT}$ & Real-Time Market price\\
\textcolor{orange}{$\Delta EV_i$}& Real Time Charging rate of PEV  $i$ deviation from DA schedule\\
\textcolor{orange}{$\Delta G_i$}& Real Time Power imported from the grid for prosumer $i$ deviation from DA schedule\\
$C_{DA}$ & DA covariance matrix for DAM price prediction error, $\in \mathbb{R}^{T \times T}$\\
$C_{RT}$ & RT covariance matrix for RTM price prediction error, $\in \mathbb{R}^{4T_H \times 4T_H}$\\
\hline \hline
\end{tabular}
\end{center}
\end{table}

We also introduce the following aggregated variables: $\textcolor{orange}{G}=\sum_{i=1}^N \textcolor{orange}{G_i}$, $\textcolor{orange}{EV}=\sum_{i=1}^N \textcolor{orange}{EV_i}$, $\textcolor{orange}{\Delta G}=\sum_{i=1}^N  \textcolor{orange}{\Delta G_i}$, $\textcolor{orange}{\Delta EV}=\sum_{i=1}^N \textcolor{orange}{\Delta EV_i}$, keeping in mind that all of these variables are $\in \mathbb{R}^T$. These equalities will be considered as constraints in the following optimization schemes. 

\section{Technical Description}

We do not present results from the Day Ahead Market in CAISO in this report, for more information on results from that market it is recommended to refer to \cite{Travacca}.

\subsection{Description of the Real Time Market in CAISO}
The real-time market in California (CAISO) has multiple market instruments in which generators can participate. There are three types of generators: (1) internal generators, ones that operate inside of the CAISO balancing area (BA) authority; (2) import and export generators, those that produce power in the BA but for use outside of the BA or those that produce power outside the BA but for use inside the BA; and, (3) dynamic resources, those generators located outside the BA that have telemetry and controls for power delivery inside the BA.\\  
\\
The two time periods that are of most relevance are the 15-minute market and the 5-minute market.  Supply-side bids are submitted, at the latest, 75 minutes prior to the start Trading Hour (T-75).  For each hour there are four bids submitted, one for each 15-minute period.  For generators that are able to participate in the 5-minute market (i.e. generators internal to the BA and dynamic resources)\footnote{With the introduction of the Energy Imbalance Market (EIM), there are now generators that participate in the real-time market that are located outside of CAISO, i.e. type 2 generators and type 3 generators.  The majority of these generators can only submit bids for the 15-minute market (type 2 generators), but a subset also participates in the 5-minute market (type 3 generators).}, the bids that are submitted for the 15-minute period will apply to the 5-minute periods within that same time period.  So a generator will provide four bids per hour and there will be three prices for each bid, thus 12 prices per hour.\\  
\\
The two real-time markets (i.e. 15-minute and the 5-minute) operate with the use of separate optimizations.  The market prices are published 45 minutes and 22.5 minutes prior to close of the 15-minute and 5-minute markets, respectively.  Those prices are published on a rolling basis, so for each 15 or 5 minute period it is 45 minutes and 22.5 minutes prior to the start of that trading period.\\
\\
It is important to note that real time markets can be positive and negative, but on average, we should expect: $\mathbb{E}(p_{RT})=p_{DA}$. Nevertheless, in practice we observe, $\mathbb{E}(p_{RT})\sim 0.93 p_{DA}$. Based on conversations with CAISO management, this is explained by some renewable generators schedule their production in real time which has a tendency to lower prices (because they have zero marginal cost). Since not all resources participate in both the DAM and RTM, there being a price difference is not surprising especially considering that the many of the low cost renewable resources are self scheduled and participate only in the RTM.  Nevertheless, while empirically there exists a price difference between the two markets, we can discard this by considering the following conditions to be true for any given time period:

\begin{itemize}
\item $p_{RT}>p_{DA} \Rightarrow$ electricity \textbf{System is Short}: day ahead supplied schedule is lower than real time demand.

\item $p_{RT}<p_{DA} \Rightarrow$ electricity \textbf{System is Long}: too much supply was scheduled compared to real time demand
\end{itemize}

The prices are received through a program called the California Market Results Interface. Figure \ref{fig:RTM} displays the Real Time Market timeline.  Two hours are depicted (hour h and hour h+1) for illustrative purposes to indicate that the the bids are rolling.  For both hours, the day-ahead market bids are shown with dotted lines and labeled as "DAM Bid".  The real-time (15-minute) bids are submitted at T-75 and the results are published at T-45.  Those bids, which were accepted are then depicted with solid lines.  The difference, $\Delta$, between the DAM and RTM is also shown.   

\begin{figure}[H]
\centering
  \includegraphics[scale=0.37]{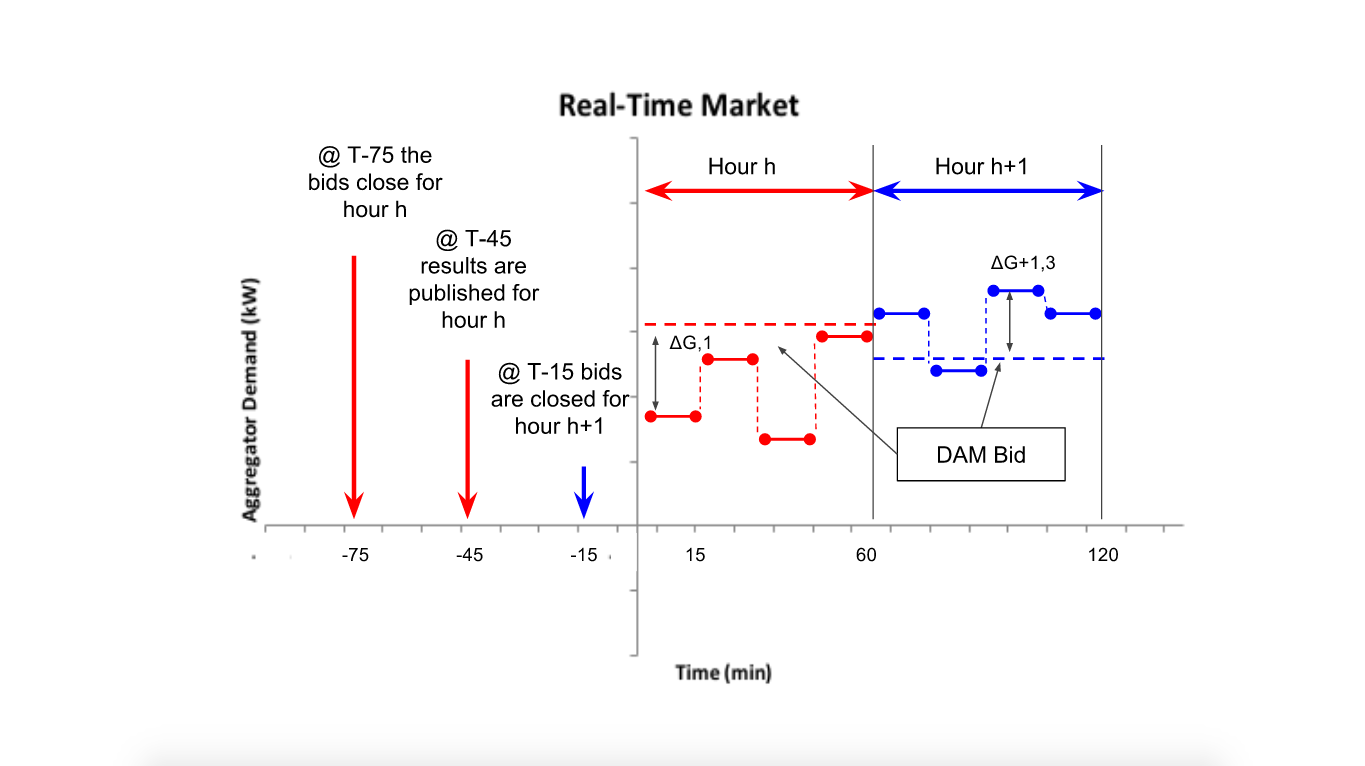}
  \caption{Graphical Depiction of the Real Time Market Process}
    \label{fig:RTM}
\end{figure}

\subsection{Market Hypothesis}
Only the 15 min market interval will be considered in this project. Note that the Smart Meters in California allow to give information every 30 min. Therefore, it is already uncertain how the settlement will be made with CAISO (i.e. proof that what was said to be consumed and produced was indeed).\\
\\
While there is a good likelihood that CAISO does require its own metering system, separate from Smart Meters, to participate in the markets in order to verify performance before any settlement.  In any case, this would not fundamentally alter the methodology taken in this project, as an aggregator would want to optimize both the 15-minute market and the 5-minute markets provided it can participate it both.

\subsection{Relation between the DAM and the Real Time Market}
At first, it seems to be a good idea to create an interaction between the Real time Market and the DAM. Indeed, as an aggregator, one can increase their revenue by predicting what is going to happen in the real time market. This consists in betting on the fact that the CAISO forecast of CAISO demand is either short or long for a given hour. Nevertheless we argue that in the CAISO market, doing so is the role of Virtual Bidding. Virtual bids must be virtual, i.e. not associated with physical bids, and the practice of using physical assets to make virtual bids is forbidden by CAISO. Therefore, we argue that the DAM and the real time market optimization should, or must, be independent: the real time market must take the results of the DAM as an exogenous input. This does not mean that the real time market does not give insight into the way to deal with the DAM. On the contrary, it can help to tune parameters and understand which constraints are important and should be kept in the DAM scheduling optimization.\\ 
\\
Moreover, doing so with physical assets would consists in exerting Market Power. All bids are assessed for whether market power is being exerted, whether intentional or not, through a Market Power Mitigation (MPM) process.  The MPM occurs prior to the optimization.  In the case that bids are deemed to be an exertion of market power, then those bids are rejected and the optimization is run without those bids.

\subsection{Price prediction and estimation of covariance matrix}

As shown in Figure 2, price fluctuation of the RTM (right figure) is much larger than that of the DAM (left figure). The DAM price was always between 0 USD/MWh and 100 USD/MWh throughout the year. In contrast, the RTM price spikes relatively frequently and went up to 1,000 USD/MWh and down to -150 USD/MWh. \\
\\

\begin{figure}[H]
\centering
  \includegraphics[scale=0.37]{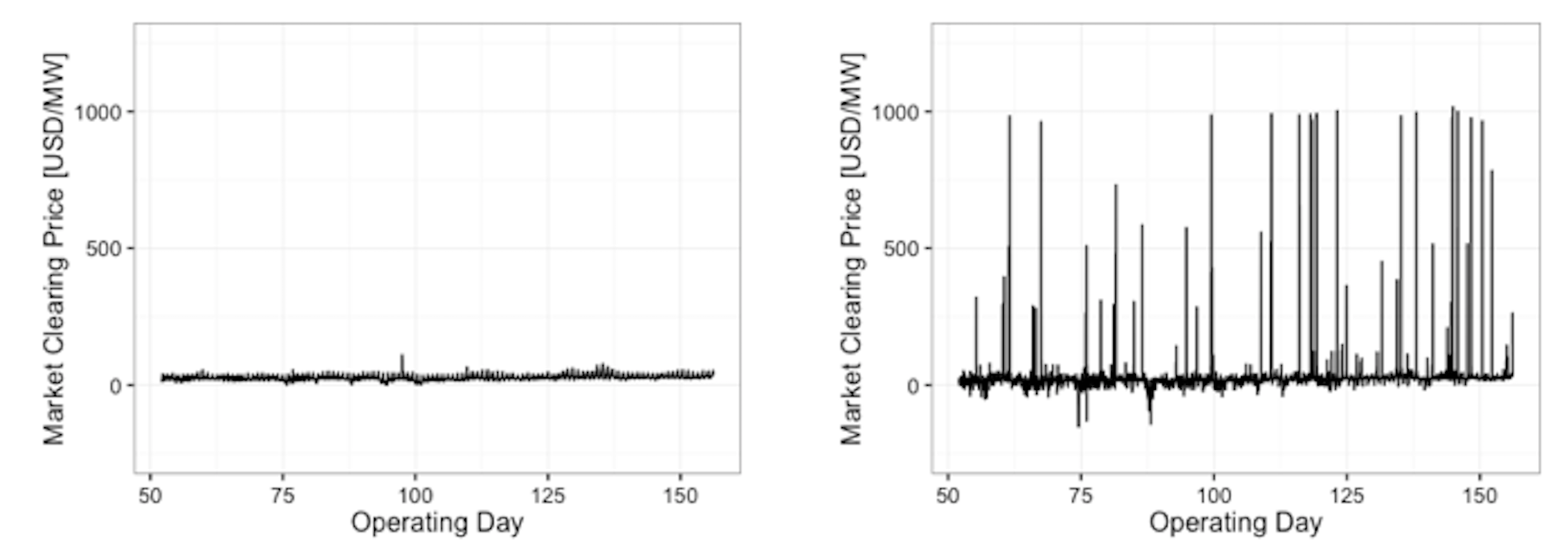}
  \caption{Comparison of price fluctuations between the Day-ahead Market and the Real-time Market}
\end{figure}

In the RTM, generation capacity tends to be constrained and supply curve (Short-term Marginal Cost: STMC) of electricity becomes more vertical (more inelastic). As a result, a small change in demand could cause price spikes in either a positive and a negative direction.  Such situations are shown in Figure 3. The figure on the left shows a change where the supply is relatively elastic and thus the price difference between $P_{RT}$ and $P_{DA}$ is small.  In contrast to the figure on the right, where the demand moves from the elastic portion of the supply curve to the inelastic portion of the supply curve and the price difference between the DAM and RTM spikes.  Nevertheless, all the questions are not answered as for now: as the price in RT are 7\% lower on average, it does seem to be a good idea for an aggregator to buy in RT only. 

\begin{figure}[H]
\centering
  \includegraphics[scale=0.37]{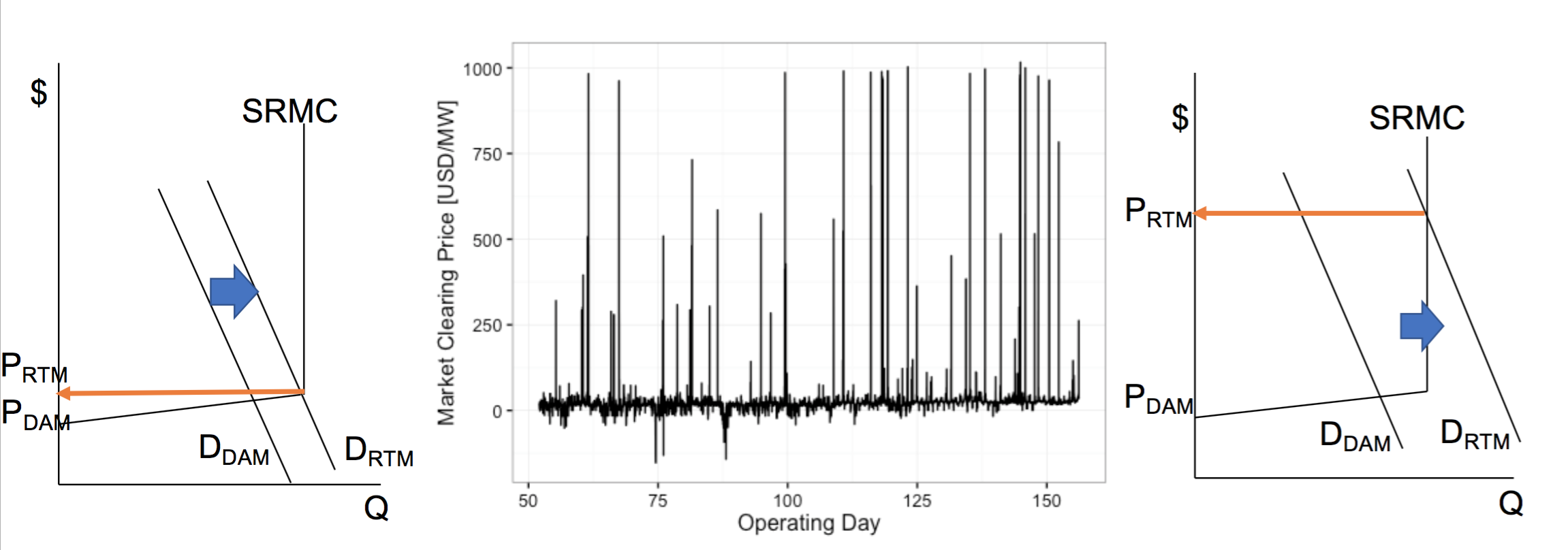}
  \caption{Mechanism of price fluctuation in the RTM}
  \label{fig:RTM}
\end{figure}

This price fluctuation of the RTM price poses a challenge in price prediction and also requires caution in dealing with the prediction uncertainty in the following step. In the stochastic optimization, we not only minimize the cost but also incorporate the uncertainty of the price prediction in the objective function. We estimate the covariance matrix that represents the uncertainty of the prediction.\\
\\
We used random forest regression to predict both the DAM and RTM prices.\\ 
\\
For the DAM price prediction, we used year, date, and hour as features (regressors). In addition to these features for the real-time forecast we use, DAM demand forecast, DAM prices, and RTM demand forecast, and operating intervals (0 min, 15 min, 30 min, and 45 min). \\
\\
We assumed multivariate normal distribution around the expected price both for RTM and DAM price prediction:\\
$$p \mathtt{\sim} N(\hat{p}, C)$$
Zero mean, normally distributed prediction error in day d is denoted as $\epsilon_d\in\mathbb{R}^{24}$. Mean squared error is then denoted as:\\
$$ \mathrm{MSE} = \frac{1}{N}\Sigma_{k=1}^{N_d}\epsilon_k \epsilon_k^T $$
It revealed to be difficult to produce an online prediction model for RTM prices, therefore we kept a 'static' one for RTM prediction. For coherence we did the same for DAM price prediction. We used the same data that we train to predict (which is not the best thing to do in practice, because it leads to over fitting, nevertheless random forest is usually doing good at avoiding that). As a consequence, the prediction we get can be seen as the best prediction we could probably get in practice: for DAM prediction we get a RMSE of 2\$ whereas we get 15\$ RMSE for the RTM.  
Given electricity price data, we can estimate the concave precision matrix using classic maximum likelihood estimation. This covariance matrices represent price prediction uncertainty and is used as an input for the following stochastic optimization model to incorporate risk aversion characteristics. We incorporate risk management using the classic Markovitz Portfolio Optimization setting (cf. following section). 

\textit{Note: we used Python to create this prediction model}

\begin{figure}[H]
  \centering
  \begin{minipage}[b]{0.4\textwidth}
    \includegraphics[width=1.3\textwidth]{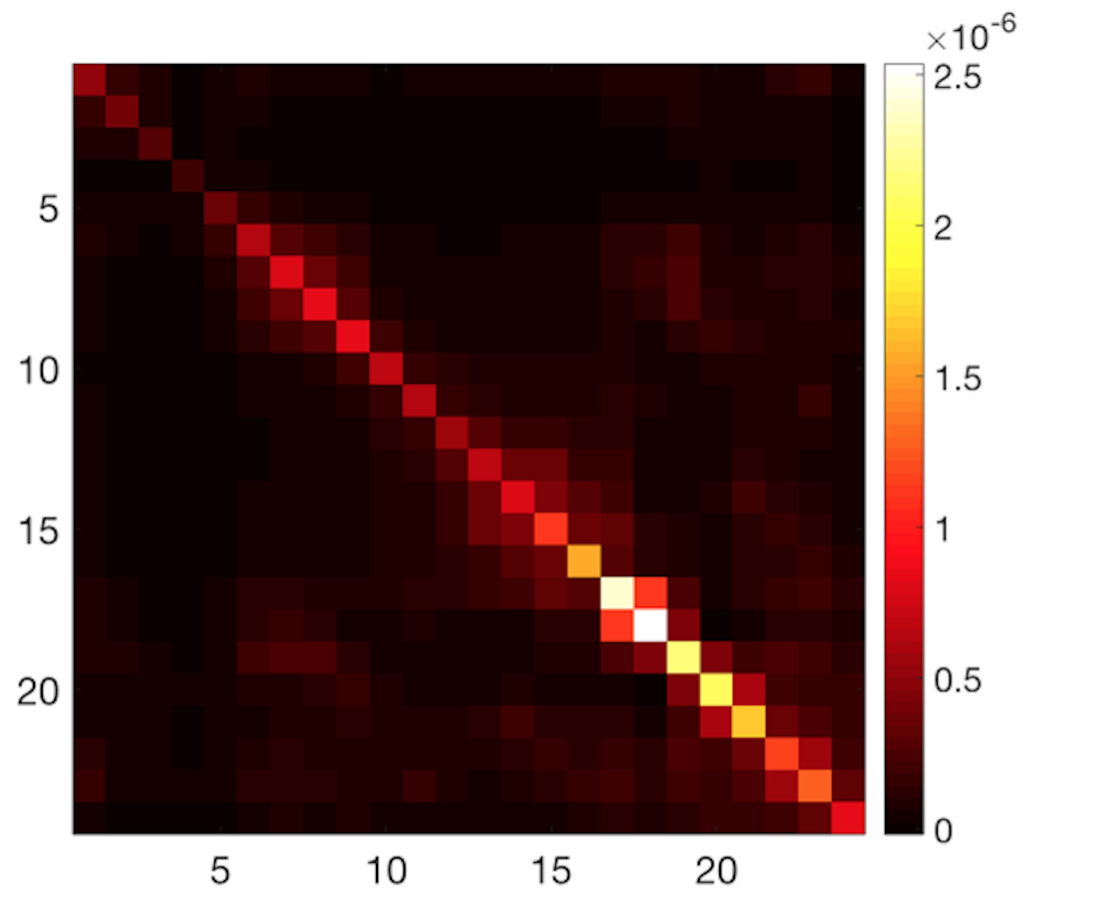}
    \caption{Covariance matrix Heatmap for DA market, $\in \mathbb{R}^{T times T}$}
  \end{minipage}
  \hfill
  \begin{minipage}[b]{0.4\textwidth}
    \includegraphics[width=1.3\textwidth]{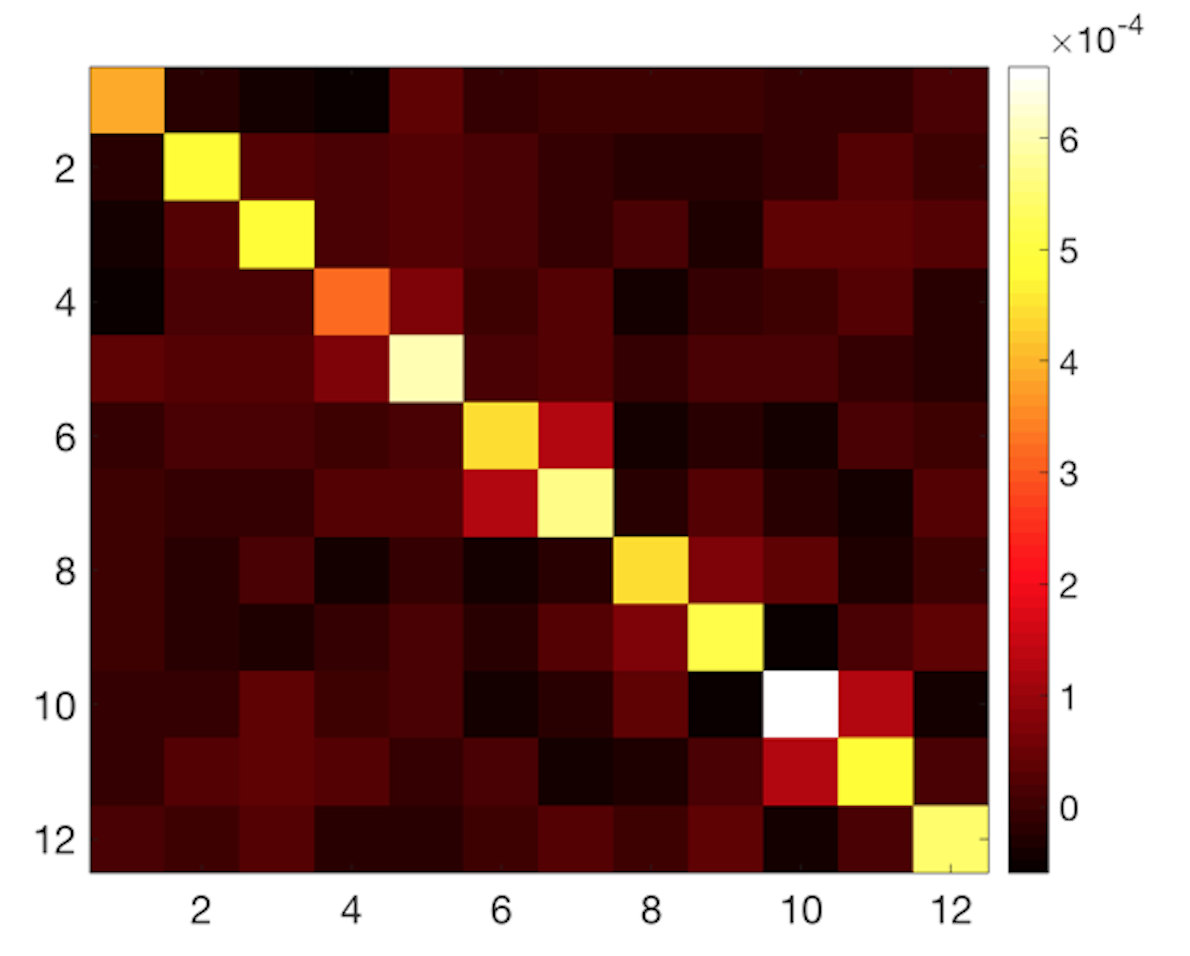}
    \caption{Covariance matrix Heatmpap for RT market, $\in \mathbb{R}^{T_H times T_H}$}
  \end{minipage}
\end{figure}

\subsection{Day-Ahead Market and Real-Time Market Optimization Objectives}
$J_{DA}$ and $J_{RT}$ denote the DA and RT objectives respectively. The choice of the quantity \textcolor{orange}{$G$} or \textcolor{orange}{$\Delta G$} can be considered as a portfolio problem where the assets is the flexibility, the returns are the DAM prices, and the budget constraint is the flexibility constraints \cite{bible}. The choice of the portfolio $G$ involves a trade-off between the expected price and the corresponding variance. For the Day-Ahead:

\begin{equation}
  J_{DAM} = 
    \underbrace{p_{DA}^T \textcolor{orange}{G}}_\text{DA total cost} + 
    \overbrace{\frac{\lambda_{DA}}{2} \textcolor{orange}{G}^TC_{DA}\textcolor{orange}{G}}^{\text{DA Risk}} 
\label{eq:DAM_objective}
\end{equation}
Now, for the RT at hour $h$, with no penalties for deviations increasing system shortness or longness, similarly:
\begin{equation}
  J_{RT}^h = 
    \underbrace{p_{RT}(h:h+4T_H-1)^T \textcolor{orange}{\Delta G}}_\text{RT total cost} + 
    \overbrace{\frac{\lambda_{RT}}{2} \textcolor{orange}{\Delta G}^TC_{RT}\textcolor{orange}{\Delta G}}^{\text{RT Risk}} 
\label{eq:RTM_objective}
\end{equation}

If the objective function for DAM is easy to understand, the Real Time objective is a little more subtle (erratum with respect to the presentation given in class, mistake that nevertheless inspired what will follow 
). Let us take a concrete example to understand why (\ref{eq:RTM_objective}) is the right objective function. This example is worked out in table \ref{rtm_ex}.

\begin{table}[H]
\caption{Why this cost function for RTM?}
\label{rtm_ex}
\begin{center}
\begin{tabular}{|p{3cm}| p{6.5cm}|p{6.5cm}|}
\hline \hline
 & $\Delta G=+1MW \geq 0$ & $\Delta G=-1MW \leq 0$\\
 \hline
 $p_{RT}=40>p_{DA}=30 \$/MW$ (SYSTEM SHORT)& $\Delta G$ increases the strain on the system. The aggregator has to pay for this extra quantity at RT market price: 30 \$/MW. Had this extra demand been scheduled in the DA, the cost would have been 10 \$/MW lower& $\Delta G$ decreases the strain on the system. The aggregator is paid as if it was providing 1MW of power supply: \$40. Had this demand reduction been included in the DA, the aggregator would have had a cost \$10 higher  \\
\hline
$p_{RT}=20<p_{DA}=30 \$/MW$ (SYSTEM LONG)& $\Delta G$ decreases the strain on the system. The aggregator has to pay for this extra quantity at RT market price: 30 \$/MW. Had this extra demand been scheduled in the DA, the cost would have been 10 \$/MW higher& $\Delta G$ increases the strain on the system. The aggregator is paid as if it was providing 1MW of power supply: \$20. Had this demand reduction been included in the DA, the aggregator would have had a \$10 lower cost  \\
\hline
$p_{RT}=-10>p_{DA}=30 \$/MW$ (SYSTEM ULTRA LONG)& $\Delta G$ decreases the strain on the system. The aggregator is paid for extra quantity at RT market price: -30 \$/MW! Had this extra demand been scheduled in the DA, the cost would have been 40 \$/MW higher& $\Delta G$ increases the strain on the system. The aggregator has to pay as if was providing 1MW of power supply: \$40. Had this demand reduction been included in the DA, the aggregator would have had a cost \$40 lower!  \\

\hline \hline
\end{tabular}
\end{center}
\end{table}

Let us stop being 'california-centric' for a minute. In Germany and the UK, the aggregator would have to face constant imbalance prices. Let $\delta_+$ be the positive imbalance price and $\delta_-$ be the negative imbalance price: the aggregator is paid $\delta_+$ for beneficial deviations (e.g. system long and $\Delta G \geq 0$) and has to pay $\delta_-$ for bad deviation (e.g. system long and $\Delta G \leq 0$). In the UK, the Imbalance market is symmetric (i.e. $\delta_-=\delta_+$), whereas in Germany $\delta_+<\delta_-$. Let us introduce this specificities in our Real-Time objective function for which there are four cases to consider:\\
\textbf{1) System Short and $\Delta G \geq 0$:}
$$\text{Imbalance Cost}_1=\delta_-\frac{sgn(p_{RT}-p_{DA})+1}{2} \odot \max\{\Delta G,0\}$$
\textbf{2) System Short and $\Delta G \leq 0$:}
$$\text{Imbalance Cost}_2=-\delta_+\frac{sgn(p_{RT}-p_{DA})+1}{2} \odot \max\{-\Delta G,0\}$$
\textbf{3) System Long and $\Delta G \geq 0$}
$$\text{Imbalance Cost}_3=\delta_+\frac{sgn(p_{RT}-p_{DA})-1}{2} \odot \max\{\Delta G,0\}$$
\textbf{4) System Long and $\Delta G \leq 0$}
$$\text{Imbalance Cost}_4=-\delta_-\frac{sgn(p_{RT}-p_{DA})-1}{2} \odot \max\{-\Delta G,0\}$$
To wrap ones mind around these formulas, reading the worked out example from table \ref{rtm_ex} is useful. We add all of these imbalance costs (1-4) to the RT objective function described by equation (\ref{eq:RTM_objective}). The new objective function is still convex (but is no longer quadratic convex), indeed $X\in \mathbb{R}^d \to \max(X,0)$ is a convex function, and convexity is preserved trough affine composition: $X\in \mathbb{R}^d \to \max(CX+D,0)$ is convex,  with $C,D$ matrices of appropriate sizes. There is nevertheless a way to transform the objective to make it a QP as described in the presentation, but we find this approach more elegant as it does not require to add new decision variables to the problem and deal with a quadratic non convex equality constraint (although we can show that it can be removed).

\subsection{Local Model for Prosumers}
Here we only display the local model for the real time model. The DAM model is thoroughly described in \cite{Travacca}, or can be understood regardless with what follows. Let $i \in \{1,...N\}$. We denote $EV_i^*$ and $G_i^*$ the optimal  schedules from DA. In the real time optimization, these variables are considered as exogenous inputs. 
\subsubsection{Local Power Balance}
The power balance (\ref{eq:powerbalance}) establishes the link between the control variables $\textcolor{orange}{\Delta EV_i}$ and $\textcolor{orange}{\Delta G_i}$.
\begin{equation}
L_i+EV_i^*+ \textcolor{orange}{\Delta EV_i}= S_i+G_i^* + \textcolor{orange}{\Delta G_i}
\label{eq:powerbalance}
\end{equation}

\subsubsection{Local Grid Constraints}
At any given node in the distribution network, there is a limit on power import or export (\ref{eq:localgrid}). Typically, for residential customers, $\overline{G}_i\simeq 10 \, kW$. 
\begin{equation}
\underline{G}_i\leq G_i^*+ \textcolor{orange}{\Delta G_i}\leq \overline{G}_i
\label{eq:localgrid}
\end{equation}

\subsubsection{Local PEV constraints}
Here, a model equivalent to the one developed in  \cite{EVmodel} is used. The PEV dynamics and state of energy constraints can be summarized as follow:
\begin{equation}
\underline{ev}_i \leq \Delta t A \cdot (EV_i^*+\textcolor{orange}{\Delta EV_i}) \leq  \overline{ev}_i 
\label{eq:evdynamic}
\end{equation}
With $A$, trigonal inferior matrix with ones. For more information please refer to \cite{Travacca}. Finally, the PEVs charging power constraint is given by (\ref{eq:evdynamic2}).
\begin{equation}
\underline{EV}_i\leq EV_i^* +\textcolor{orange}{\Delta EV_i} \leq \overline{EV}_i
\label{eq:evdynamic2}
\end{equation}

Note that when the PEV is un-plugged at a given time $t \in \{1,...T\}$, then  $\underline{EV}_i^t=\overline{EV}_i^t=0$ is required. \textbf{For the purpose of conciseness, the set of local constraints (\ref{eq:powerbalance}), (\ref{eq:localgrid}), (\ref{eq:evdynamic}) and (\ref{eq:evdynamic2}) are hereby referred to as $local_i$. }

\subsection{Model Predictive Control Scheme Scheme for Real-Time Operation}
\begin{algorithm}
\caption{Model Predictive Control}\label{euclid}
\begin{algorithmic}[1]
\State \textbf{Initialization}: No warm start possible with CVX
\For {$h$ from $1$ to $T$ do:}
$$\Delta EV^*, \Delta G^*=\argmin{J_{RT}^h(\Delta EV, \Delta G,\Delta EV_i, \Delta G_i )} \text{ s.t. local}_i $$
Only implement the decision for the first hour: 
$\Delta EV^*=\Delta EV^*(1:4)$, $\Delta G^*=\Delta G^*(1:4)$, etc. 
\EndFor
\State \textbf{End For}
\end{algorithmic}
\end{algorithm}

\textit{Note: as stated by Michael C.Grant, one of the creators of CVX, it is not possible to impose an initial guess onto CVX. Therefore creating our own solver in the future is of particular interest (or using quadprog in Matlab). We can see here the limitations to CVX. }
\section{Data}
For price prediction and covariance matrix estimation, we collected CAISO's DAM (hourly) and RTM (15 minutes) price and demand forecast data in PG\&E area from January 2013 to March 2017.\\
\\
For stochastic optimization, we used high resolution data in San Francisco. We used mobility data  from 2,000 full electric vehicle's in the bay area. A single solar PV generation data was generated using PVsim (SunPower). The load was modeled taking aggregate load data from CAISO in the PG\&E region and adding random noise to it to create heterogeneity between prosumers (idem for prosumer PV production). Most of the data was concatenated in the same CSV file using R.   

\section{Simulation and Results}
\textit{Note: the code can be made available upon request (Python program, Matlab scripts and files, R program for data cleaning)}\\
We used Matlab to illustrate our method and theory.
Given the time frame, it was difficult to realize more than a one day simulation: the main reason is the fact that the way the constraint bounds (eq. \ref{eq:evdynamic2}) are generated from data is complex, and it was tailored for one day simulations only. Nevertheless, this one day simulation allows us to see the behavior of our scheduling and real time operation method. We made a simulation for 100 prosumers, the parameters that were chosen for this simulation are provided in table \ref{parameters}.

\renewcommand{\arraystretch}{1.8}
\begin{table}[H]
\caption{Parameter Values}
\label{parameters}
\begin{center}
\begin{tabular}{c l}
\hline \hline
$N$ & 100 \\
$T_H$ & 3 hours\\
$\overline{G}$ & 10 kW\\
$\eta$ & 90\% (charging efficiency)\\
$delta_-,delta_+$ & 0\\
$\lambda_{DA},\lambda_{RT}$ & 1 \\
\hline \hline
\end{tabular}
\end{center}
\end{table}

\begin{figure}[h]
  \centering
  \begin{minipage}[b]{0.4\textwidth}
    \includegraphics[width=1.4\textwidth]{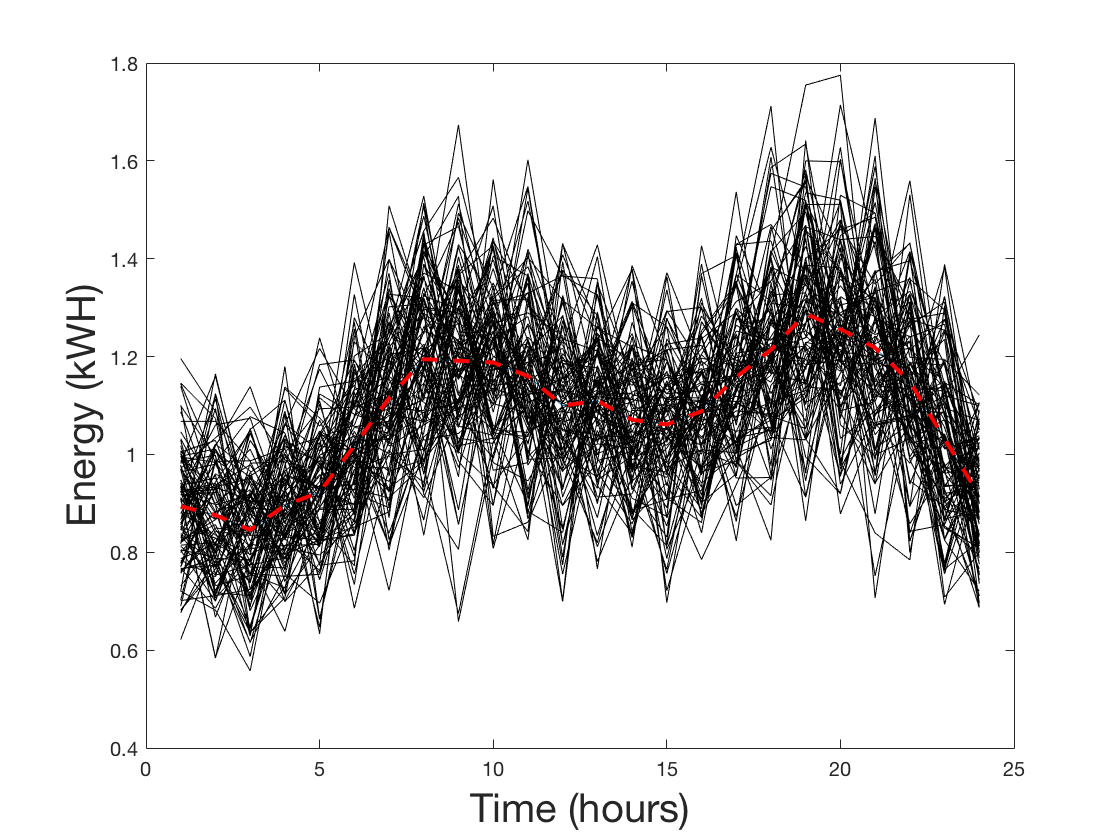}
    \caption{Local uncontrollable load consumptions (black curves), Average (red dashed curve)}
    \label{local_load}
  \end{minipage}
  \hfill
  \begin{minipage}[b]{0.4\textwidth}
    \includegraphics[width=1.4\textwidth]{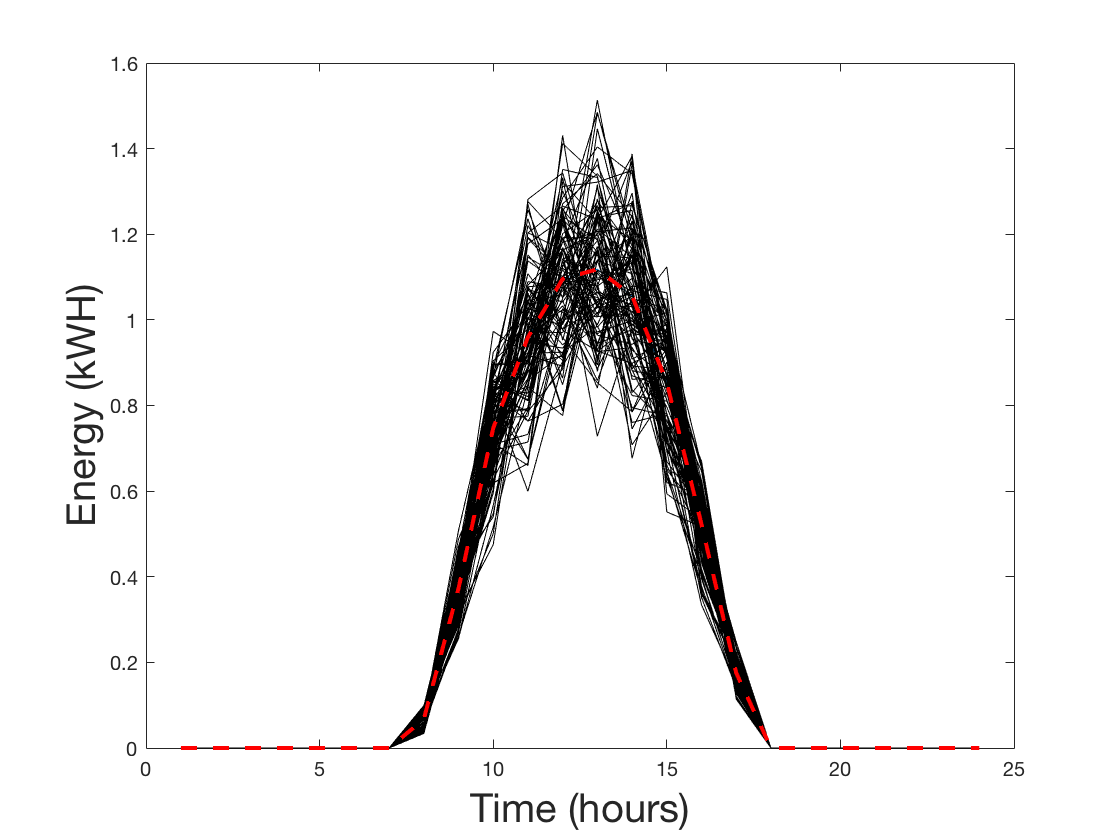}
    \caption{Local solar PV production (black curves), Average solar PV production (red dashed curve)}
        \label{local_PV}
  \end{minipage}
\end{figure}

\begin{figure}[H]
  \centering
  \begin{minipage}[b]{0.4\textwidth}
    \includegraphics[width=1.4\textwidth]{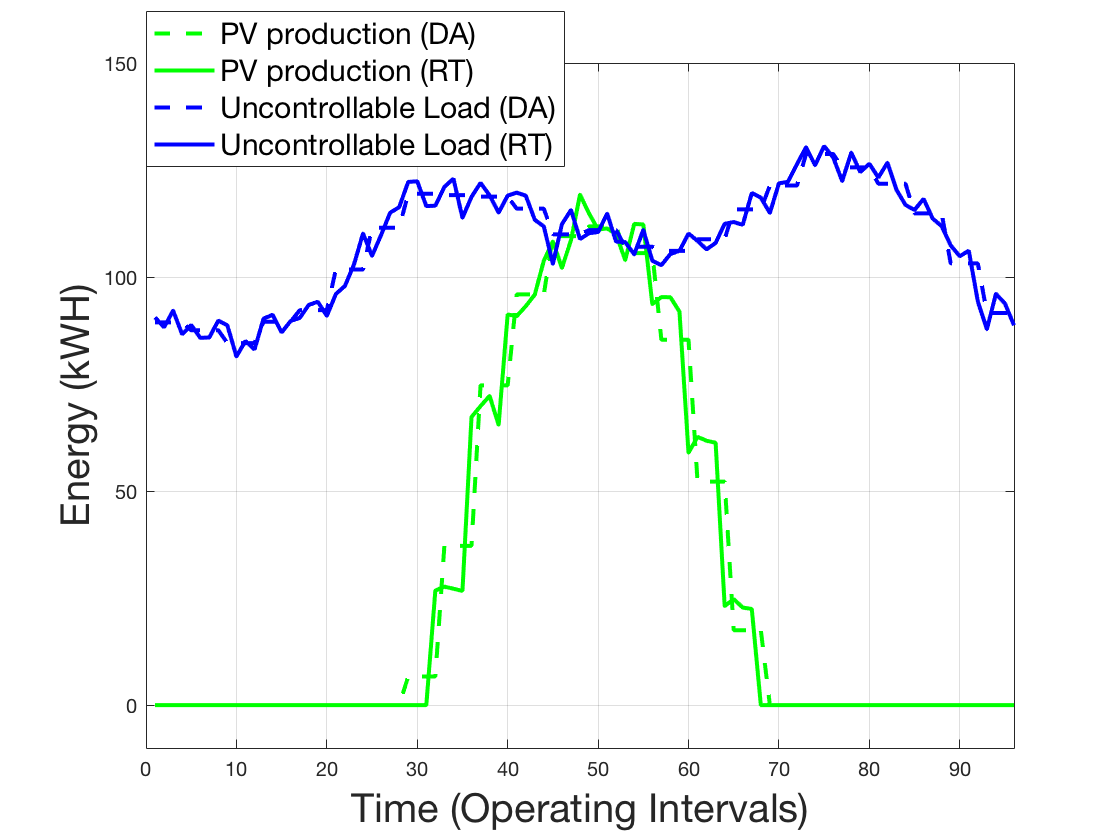}
    \caption{Aggregated uncontrollable load consumption and solar production: difference between RT and DA}
    \label{aggregate_load_PV}
  \end{minipage}
  \hfill
  \begin{minipage}[b]{0.4\textwidth}
    \includegraphics[width=1.4\textwidth]{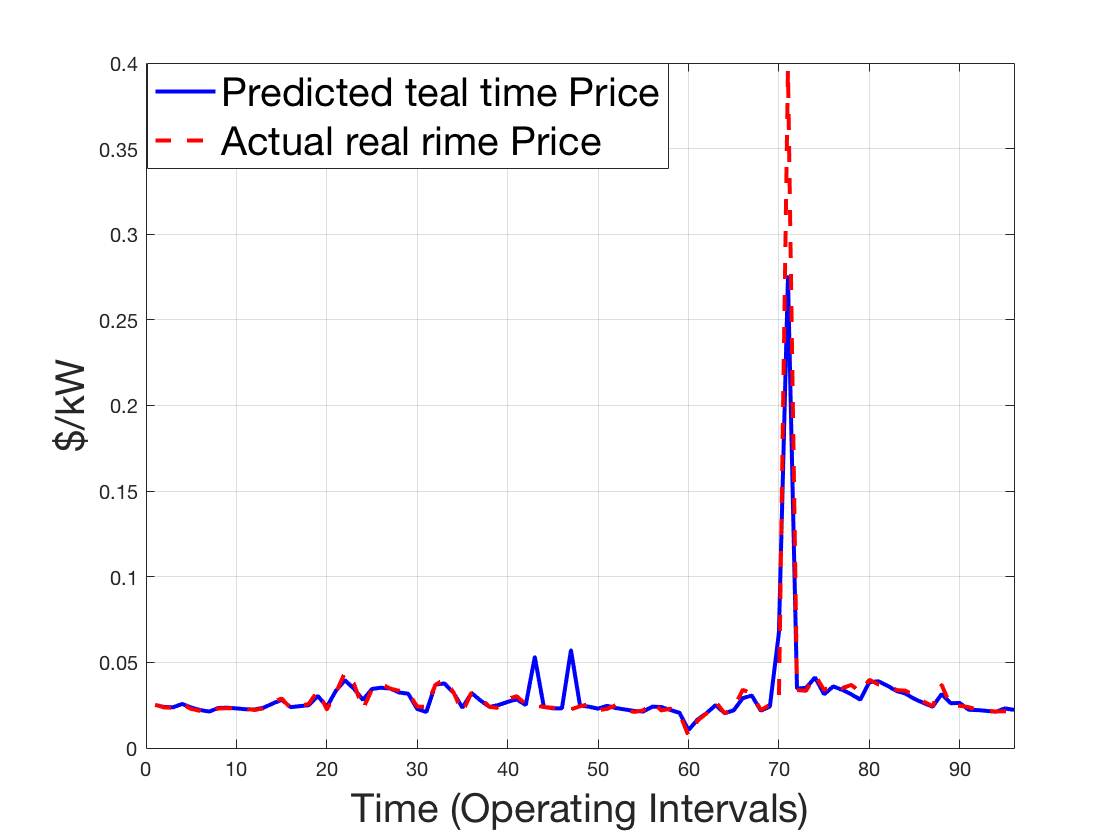}
    \caption{Real-Time prices for the chosen day of simulation}
    \label{RTprice}
  \end{minipage}
\end{figure}

\begin{figure}[H]
\centering
  \includegraphics[scale=0.18]{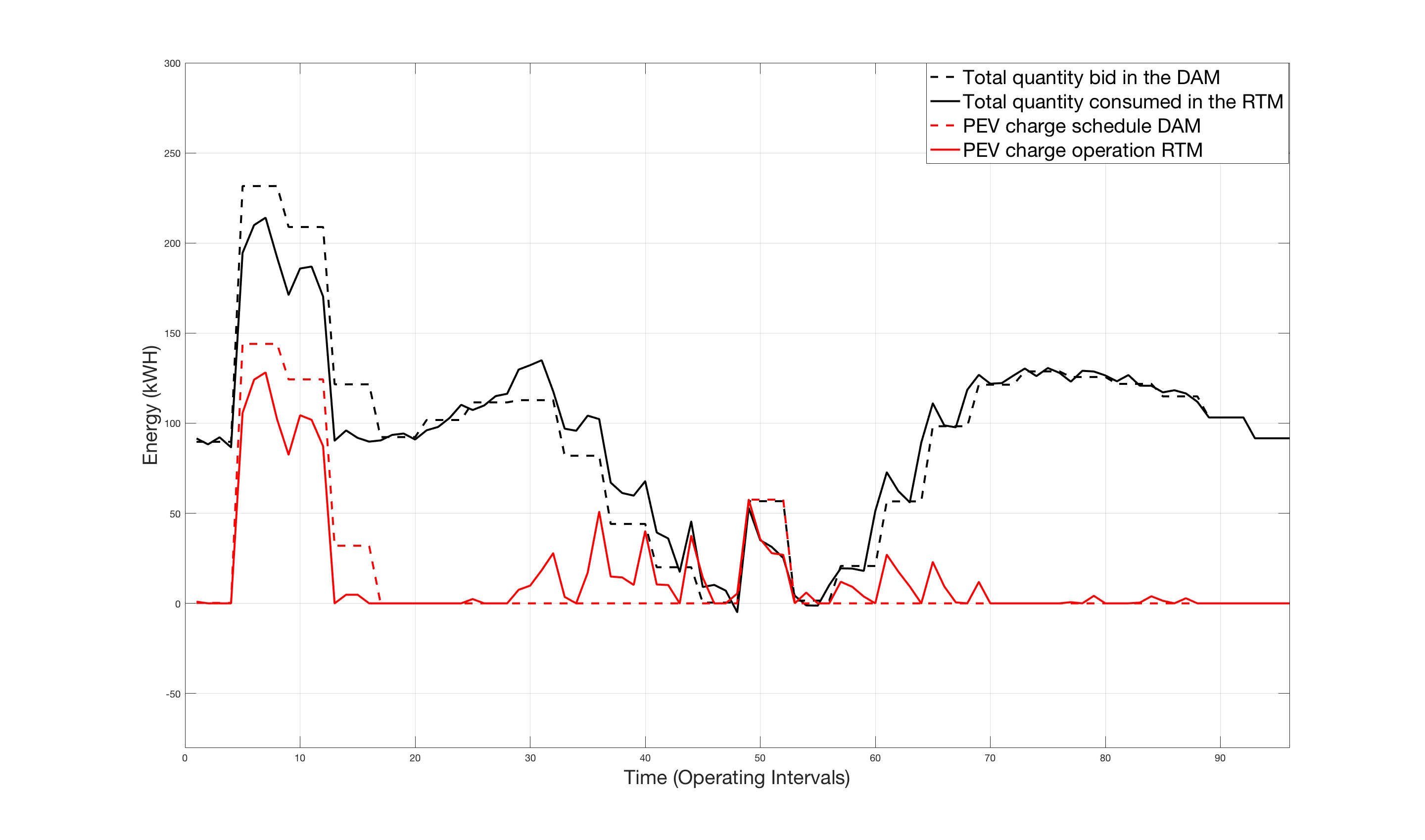}
  \caption{DA Schedule and RT operation results: $G^*$, black dashed curve, $\Delta G^*$ solid black dashed curve, $EV^*$ red dashed curve and $\Delta EV^*$ solid red curve}
  \label{results}
\end{figure}

First, figures \ref{local_load} and \ref{local_PV} provide a visualization of the average uncontrollable load and PV production as well as the heterogeneity inside the pool for DA (i.e. 1 hour time step). Figure \ref{aggregate_load_PV} provides a visualization of the difference between the DA and RT aggregated load an PV production. This difference is only known in the RT operation process. Finally, Figure \ref{RTprice} shows the real time price for the day. The results are provided in figure \ref{results}. This figure shows the difference between the schedules and real time operation. In the DA we predict that our clearing price will be \$68.7 for the next day. In reality, when the DA market clears, the aggregator realizes that its cost will be \$71.1. In the RTM, the total supplementary cost over the day is \$0.30. This means that the aggregator is not able to leverage any value in the RT market. Nevertheless, our scheme gives us a way to do real time operation over our pool. It is difficult to determine what costs the aggregator would incur for its real time operation without this scheme, moreover the price spike only appears for 15 min.  Based on our simulations, the aggregator does not take advantage of this price surge. This is due to the fact that our MPC horizon is too small and there is no available flexibility by the time we reach spike time (around operating interval 70). When the risk aversion parameter for Real Time $\lambda_{RT}$ is set to zero, the total cost is -\$0.90, which means the aggregator leveraged the RTM to reduce its cost. Nevertheless, based on the simulation, the aggregator is still not able to take advantage of the surge price. 

\section{Summary and Future Works}
This project involved advancing existing research, through developing a forecast model for the day-ahead market and real-time markets using random forest regression technique (although not the primary objective), we then propose a two-step method for managing a pool of prosumers with local production and flexibility. In a first step, the aggregator schedules its total load to CAISO, and in a second step, the aggregator operates the pool in the RTM using a Model Predictive Control Technique to decide how it should deviate from the DAM schedule.\\
\\
A lot of effort was put in producing a model and a method to tackle scheduling and operations for DERs and flexibility. Nevertheless, a large number of questions remain unanswered and a lot is yet to be done. First, an online prediction should be built. If random forest regression performs well for DAM price prediction, it does not seem to be the case for RT. We should dive into the vast literature of real-time surge pricing in RT market. Second, we need to produce some simulations for several days to be able to tune parameters (e.g. risk aversion and time horizon). Third, we need to test the method we developed for European real-time imbalance markets. We concede that this list is far from being exhaustive.

\end{document}